\documentclass[longauth]{aa}

\usepackage{graphicx}
\usepackage{txfonts}
\usepackage{float}
\usepackage{latexsym}
\usepackage{amssymb}
\usepackage{amsmath}
\usepackage{amsfonts}
\usepackage{caption}
\usepackage{subcaption}
\usepackage{soul}
\usepackage[hidelinks,colorlinks=true,linkcolor=blue,citecolor=blue]{hyperref}
\usepackage[utf8]{inputenc}
\usepackage[english]{babel}
\usepackage{import}
\usepackage{csvsimple}
\usepackage{booktabs}
\usepackage{natbib}
\usepackage{multirow}

\bibpunct{(}{)}{;}{a}{}{,}
\begin{document} 

   \title{The BINGO Project IX: } 
   \subtitle{Search for Fast Radio Bursts -- A Forecast for the BINGO Interferometry System\thanks{The authors dedicate this work to the memory of Frederico Augusto da Silva Vieira (1984 --2023).}}

\author{   
    Marcelo V. dos Santos\thanks{marcelo.santos@df.ufcg.edu.br}\inst{\ref{inst:ufcg}, \ref{inst:usp}}
    \and
    Ricardo G. Landim\thanks{ricardo.landim@port.ac.uk}\inst{\ref{inst:tum}, \ref{inst:icg}}
    \and
    Gabriel A. Hoerning\thanks{ghoerning@usp.br}\inst{\ref{inst:usp}}
    \and
    Filipe B. Abdalla\thanks{filipe.abdalla@gmail.com}\inst{\ref{inst:usp}, \ref{inst:inpe}, \ref{inst:ucl},\ref{inst:ru}}
    \and
    Amilcar Queiroz\inst{\ref{inst:ufcg}}
    \and
   Elcio Abdalla\inst{\ref{inst:usp}}
    \and
    Carlos A. Wuensche  \inst{\ref{inst:inpe}}
    \and
    Bin Wang\inst{\ref{inst:yzu},\ref{inst:sjtu}}
    \and
    Luciano Barosi\inst{\ref{inst:ufcg}}
    \and
    Thyrso Villela\inst{\ref{inst:inpe},\ref{inst:cgee},\ref{inst:unb}}
    \and 
    Alessandro Marins\inst{\ref{inst:ustc},\ref{inst:usp}} 
    \and
    Chang Feng \inst{\ref{inst:ustc},\ref{inst:ustc2},\ref{inst:ustc3}}
    \and
    Edmar Gurj\~ao \inst{\ref{inst:ufcg}}
    \and
    Camila P. Novaes\inst{\ref{inst:inpe}}
    \and 
    Larissa C. O. Santos\inst{\ref{inst:yzu}}
    \and
    Jo\~ao R.L. Santos  \inst{\ref{inst:ufcg}}
    \and
    Jiajun Zhang\inst{\ref{inst:shao}}
    \and
    Vincenzo Liccardo\inst{\ref{inst:inpe}}
\and
Xue Zhang  \inst{\ref{inst:yzu}}
    \and
Yu Sang  \inst{\ref{inst:yzu}}
    \and
Frederico Vieira\thanks{\emph{In memoriam}}  \inst{\ref{inst:inpe}}
\and 
Pablo Motta \inst{\ref{inst:usp}}
}

\institute{
    Unidade Acad\^emica de F\'{i}sica, Universidade Federal de Campina Grande, R. Apr\'{i}gio Veloso,  Bodocong\'o, 58429-900 - Campina Grande, PB, Brazil \label{inst:ufcg}
    \and
    Instituto de F\'{i}sica, Universidade de S\~ao Paulo, R. do Mat\~ao, 1371 - Butant\~a, 05508-09 - S\~ao Paulo, SP, Brazil \label{inst:usp}
    \and
    Technische Universit\"at M\"unchen, Physik-Department T70, James-Franck-Strasse 1, 85748, Garching, Germany\label{inst:tum}
    \and
    Institute of Cosmology and Gravitation, University of Portsmouth, Dennis Sciama Building, Portsmouth PO1 3FX, United Kingdom\label{inst:icg}
    \and
    Instituto Nacional de Pesquisas Espaciais, Av. dos Astronautas 1758, Jardim da Granja, S\~ao Jos\'e dos Campos, SP, Brazil\label{inst:inpe}
    \and
    University College London, Gower Street, London, WC1E 6BT, UK\label{inst:ucl}
    \and
    Department of Physics and Electronics, Rhodes University, PO Box 94, Grahamstown, 6140, South Africa\label{inst:ru}
    \and
    Center for Gravitation and Cosmology, Yangzhou University, 224009, China\label{inst:yzu}
    \and
    College of Science, Nanjing University of Aeronautics and Astronautics, Nanjing 211106, China\label{inst:nuaa}
    \and
    Shanghai Astronomical Observatory, Chinese Academy of Sciences, Shanghai 200030, China \label{inst:shao}
    \and
    School of Aeronautics and Astronautics, Shanghai Jiao Tong University, Shanghai 200240, China \label{inst:sjtu}
    \and
    Centro de Gest\~ao e Estudos Estrat\'egicos SCS Qd 9, Lote C, Torre C S/N Salas 401 a 405, 70308-200 - Bras\'ilia, DF, Brazil\label{inst:cgee}
    \and
    Instituto de F\'{i}sica, Universidade de Bras\'{i}lia, Campus Universit\'ario Darcy Ribeiro, 70910-900 - Bras\'{i}lia, DF, Brazil \label{inst:unb}
    \and
    Department of Astronomy, School of Physical Sciences, University of Science and Technology of China, Hefei, Anhui 230026, China\label{inst:ustc}
    \and
    CAS Key Laboratory for Research in Galaxies and Cosmology, University of Science and Technology of China, Hefei, Anhui 230026, China\label{inst:ustc2}
    \and
    School of Astronomy and Space Science, University of Science and Technology of China, Hefei, Anhui 230026, China\label{inst:ustc3}
}

\authorrunning{M.V dos Santos et al.}
\titlerunning{The BINGO Project IX: Search for Fast Radio Bursts -- A Forecast for the BINGO Interferometry System}

\date{Received: Accepted:}


\abstract
    {The \textbf{B}aryon Acoustic Oscillations (BAO) from \textbf{I}ntegrated \textbf{N}eutral \textbf{G}as \textbf{O}bservations (\textbf{BINGO}) radio telescope will use the neutral Hydrogen emission line to map the Universe in the redshift range  $0.127 \le z \le 0.449$, with the main goal of probing BAO. In addition, the instrument optical design and hardware configuration support the search for \textbf{F}ast \textbf{R}adio \textbf{B}ursts (\textbf{FRBs}).}
    {In this work, we propose the use of a \textbf{B}INGO \textbf{I}nterferometry \textbf{S}ystem (\textbf{BIS}) including new auxiliary, smaller, radio telescopes (hereafter \emph{outriggers}). The interferometric approach makes it possible to pinpoint the FRB sources in the sky. We present here the results of several BIS configurations combining BINGO horns with and without mirrors ($4$ m, $5$ m, and $6$ m) and 5, 7, 9, or 10 for single horns.}
    {We developed a new {\tt Python} package, the {\tt FRBlip}, which generates synthetic FRB mock catalogs and computes, based on a telescope model, the observed signal-to-noise ratio (S/N) that we used to compute numerically the detection rates of the telescopes and how many interferometry pairs of telescopes (\emph{baselines}) can observe an FRB. FRBs observed by more than one baseline are the ones whose location can be determined. We thus evaluate the performance of BIS regarding FRB localization.}
    {We found that BIS will be able to localize 23 FRBs yearly    with single horn outriggers in the best configuration (using 10 outriggers of 6 m mirrors), with redshift $z \leq 0.96$; the full localization capability depends on the number and the type of the outriggers. Wider beams are best to pinpoint FRB sources because potential candidates will be observed by more baselines, while narrow beams look deep in redshift. 
    }
    {The  BIS can be a powerful 
    extension of the regular BINGO telescope, dedicated
    to observe hundreds of FRBs during Phase 1. Many of them will be well localized with a
      single horn + 6 m dish as outriggers. 
   }

   \maketitle
   
  

\section{Introduction}
\label{intro}

Fast Radio Bursts (FRBs) are transient radio pulses discovered in 2007 using data from the Parkes telescope \citep{Lorimer:2007qn, Thornton2013}. Their origin is still an open problem. They are very bright, with sub-second duration and their arrival time delay is characterized by a well-known frequency dependency (see \citealt{Petroff:2019tty, Petroff:2021wug} for current reviews). There are over 830 detected FRBs at the time of this writing,\footnote{See \url{https://www.wis-tns.org/} for the official list of events.} where 206 are repeaters (which can be periodic or not) and 19 host galaxies have been localized \citep{Petroff:2021wug}, with no apparent privileged sky distribution. 

Detections have been made in observational frequencies ranging from $\sim 110$  MHz \citep{Pleunis:2020vug, Pastor-Marazuela:2020tii} to $\sim 8 $  GHz \citep{Gajjar:2018bth}.  The emitted radio waves are dispersed by the free electrons in the medium along the line-of-sight, and the usually large dispersion measures (DM), which exceeded the Milky Way limits \citep{Cordes2001, 2019ascl.soft08022Y}, initially suggested that they have only an extragalactic origin. However,  an FRB was also detected in the Milky Way \citep{CHIMEFRB:2020abu}. A number of celestial objects are being considered as FRB progenitor candidates,  including active galactic nuclei,  supernovae remnants, and mergers of compact objects such as neutron stars, black holes and white dwarfs \citep{Platts:2018hiy}.\footnote{See \url{https://frbtheorycat.org/} for a list of different models.}  

FRBs have been used as a tool for different cosmological and astrophysical studies, e.g. the fraction of baryon mass in the intergalactic medium  \citep[IGM;][]{Munoz:2018mll, Walters:2019cie, Qiang:2020vta}, dark energy \citep{Walters2018, Liu:2019jka, Zhao:2020ole}, the equivalence principle \citep{Wei:2015hwd,Tingay:2015wdv,Tingay:2016tgf, Shao2017,Yu:2017xbb,Bertolami:2017opd,Yu:2018slt,Xing:2019geq}, IGM foreground and halos \citep{Zhu:2020prt}, dark matter \citep{Munoz:2016tmg,Wang:2018ydd,Sammons:2020kyk, Landim:2020ked}, and other astrophysical problems \citep{Yu:2017beg, Wu:2020jmx, Linder:2020aru}. However, current studies relying on FRBs are still limited due to the small sample of identified hosts, which allows for their redshift determination.


Several telescopes have detected FRBs, such as  Parkes    \citep{Lorimer:2007qn}, Arecibo telescope \citep{Spitler:2014fla}, Green Bank  \citep{Masui:2015kmb},  ASKAP   \citep{Bannister:2017sie},   UTMOST \citep{Caleb:2017vbk}, CHIME  \citep{Amiri:2018qsq}, and Apertif  \citep{Connor:2020oay}, among others. CHIME has had an important role in discovering new FRBs. It detected 536 FRBs observed between 400 MHz and 800 MHz during one year \citep{CHIMEFRB:2021srp}, increasing enormously the total number of known events.

The task of determining the redshift of the host galaxy is still a hard one. Indeed, to determine the distance one should localize the events with arcsecond precision. We can use other surveys to find their optical counterpart \citep{Tendulkar2017host, Bhandari:2021pvj}. 
Localizing an FRB event is very important to understand the origin and environment of its progenitor, as well as to use its redshift determination to attack a few open problems in cosmology.
The precise localization of the events is possible using interferometric techniques, where the data from different antennas are cross-correlated, pinpointing the origin of the emission. We want to correlate the data from a main telescope with data from its outriggers, which, in this context, are auxiliary smaller radio telescopes. 


The BINGO (Baryon acoustic oscillations from Integrated Neutral Gas Observations) telescope \citep{Battye:2013, Abdalla:2021nyj,Abdalla:2020ypg} is 
a single-dish instrument, with a 40-m diameter primary reflector and mounted in a crossed-Dragone configuration, observing the sky in the frequency band 980 MHz -- 1260 MHz in transit mode. During Phase 1 (which should last 5 years), it will operate with 28 horns, each one attached to a pseudo-correlator equipped with 4 receivers. The optical design should deliver a final angular resolution of $\approx 40$ arcmin \citep{Wuensche:2021dcx, Abdalla:2021xpu}.
BINGO is being built in the northeast region of Brazil and its primary goal is to observe the 21-cm line of the atomic Hydrogen to measure Baryon Acoustic Oscillations (BAO) using the Intensity Mapping (IM) technique. Although it is designed for cosmological investigations, its 
large survey area can provide ancillary science in radio transients \citep{Abdalla:2021nyj}. 

To improve BINGO's capability to detect and localize FRBs,  outrigger units are being planned and designed.  
The instrumental requirements and capabilities of these outriggers are being determined using a combination of mock FRB catalogs, simulations of events, and detection forecasting.
The publicly available code {\tt frbpoppy} \citep{Gardenier:2019jit} generates  cosmological populations of FRBs and simulates surveys. However, it does not include several tools and distributions of physical quantities that would be necessary to explore different configurations and designs for  BINGO and  its outriggers, in order to produce reliable estimates of the number of detected and localized events.


The goal of this paper is to extend the initial analysis presented in \cite{Abdalla:2021nyj}, introducing the \textit{BINGO interferometry system} (BIS) and presenting details of the cross-correlation between BINGO and its proposed outriggers. The main setup of the BIS is explained in Section \ref{sec:bis}. 
We will introduce {\tt FRBlip}, a {\tt Python} package developed by the BINGO collaboration, that generates FRB mock catalogs. {\tt FRBlip} includes several distribution functions presented in \cite{Luo:2018tiy,Luo:2020wfx}, whose description and corresponding simulations are shown in Section \ref{sec:frblip}.
In Section \ref{sec:results} we present the different outrigger configurations, compare their performances, and discuss the different criteria to define the detection and localization of a FRB.
Section \ref{sec:conclusions} is reserved for conclusions.
 
Other aspects of the BINGO telescope, including instrument description, component separation techniques, simulations, forecasts for cosmological models, and BAO signal recoverability are found in the previous papers of the series \citep{Abdalla:2021nyj, Wuensche:2021dcx, Abdalla:2021xpu, Liccardo:2021kbu, Fornazier:2021ini, Zhang:2021tcl, Costa:2021jsk, Novaes:2022}, as well as in \cite{deMericia:2022twy} and \cite{Marins:2022gaz}.


\section{The BINGO Interferometry System}\label{sec:bis}

In this work, we suppose that radio telescopes are well characterized by only seven quantities: system temperature ($T_{\rm sys}$), forward gain ($G$), 
sensitivity constant ($K$, depends on the receiver type), number of polarizations ($n_p$), the frequency bandwidth ($\Delta\nu = \nu_2 - \nu_1$), a reference frequency ($\nu_1 < \nu_{\rm ref} < \nu_2$), and sampling time ($\tau$). An important derived quantity is the instrument noise
\begin{equation}\label{eq:Smin0}
    S_{\rm{min}}^{(0)} = \frac{ K \, T_{\rm{sys}}}{G \sqrt{n_p\Delta\nu\, \tau}}\,,
\end{equation}
known as sensitivity, 
which roughly defines the minimum detectable flux density assessed by a given instrument \citep{Kraus1986radio}.
In our opening paper \citep{Abdalla:2021nyj} we assumed the telescope beam as being a top hat function, constant inside a given radius. Now, we assume a (more realistic) Gaussian beam pattern,
\begin{equation}
    P_n(\mathbf{n}) = \exp\left(-4(\log2)\frac{\theta^2}{\theta_{1/2}^2}\right)\,,
\end{equation}
where $\theta$ is the angular separation to the beam center. Here, $\theta_{1/2}$ is the \emph{full width half maximum} (FWHM), related to $G$, and $\lambda_{\rm ref}$ ($\nu_{\rm ref}/c$) by 
\begin{equation}\label{eq:fwhm}
    \theta_{1/2} \approx \sqrt{\frac{4\log 2}{\pi k_B G}}\lambda_{\rm ref}\,.
\end{equation}
The forward gain $G$ is related to the effective area ($A_{\rm eff}$) as
\begin{equation}
    G = \frac{A_{\rm eff}}{2k_B}\,,
    \end{equation}
where $k_B$ is the Boltzmann constant.

In this case, the signal of a point source, such as FRBs, will be contaminated by instrument noise which depends on its sky position ($\mathbf{n}$) as
\begin{equation}\label{eq:Smin}
    S_{\rm{min}}(\mathbf{n}) = \frac{S_{\rm{min}}^{(0)}}{{ P_n}(\mathbf{n})}\,,
\end{equation}
which we name directional sensitivity. The normalized antenna pattern appears in the denominator as a matter of convenience, because it will appear in the numerator of Eq. (\ref{eq:snr_i}).

\begin{figure}[ht]
\centering
\includegraphics[width=0.9\columnwidth]{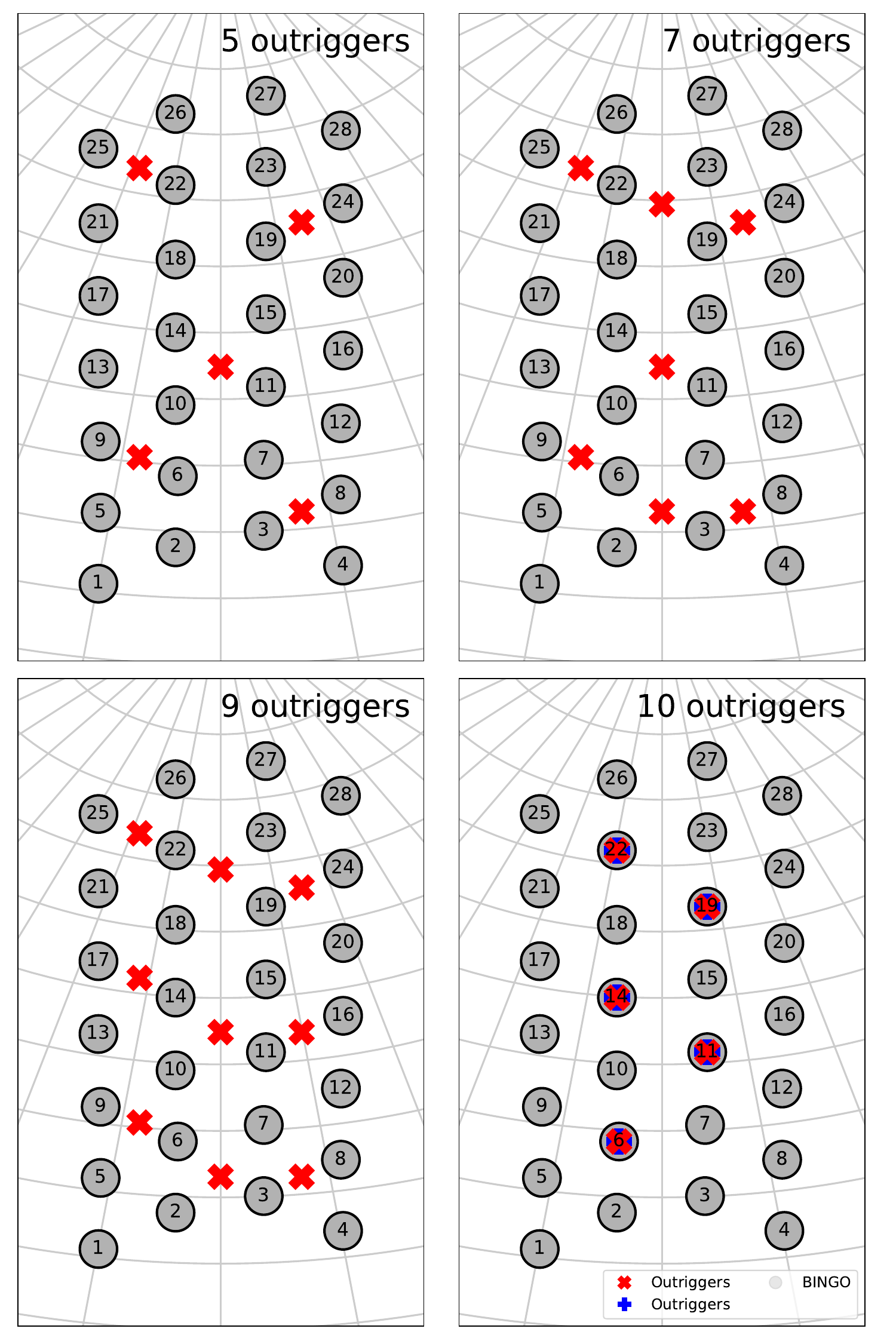}
\caption{BINGO beams on the sky and outriggers, and four different outriggers pointing. In the case of 10 outriggers, the pointings are directly overlappping  with some of the BINGO beams in order to investigate how two equal outriggers pointing in the same direction at the same time affects the the localization. The red $\times$ outriggers and the blue $+$ outriggers are equal telescopes poiting to same directions.}
\label{fig:pointings}
\end{figure}

\begin{figure}
    \centering
\includegraphics[width=0.9\columnwidth]{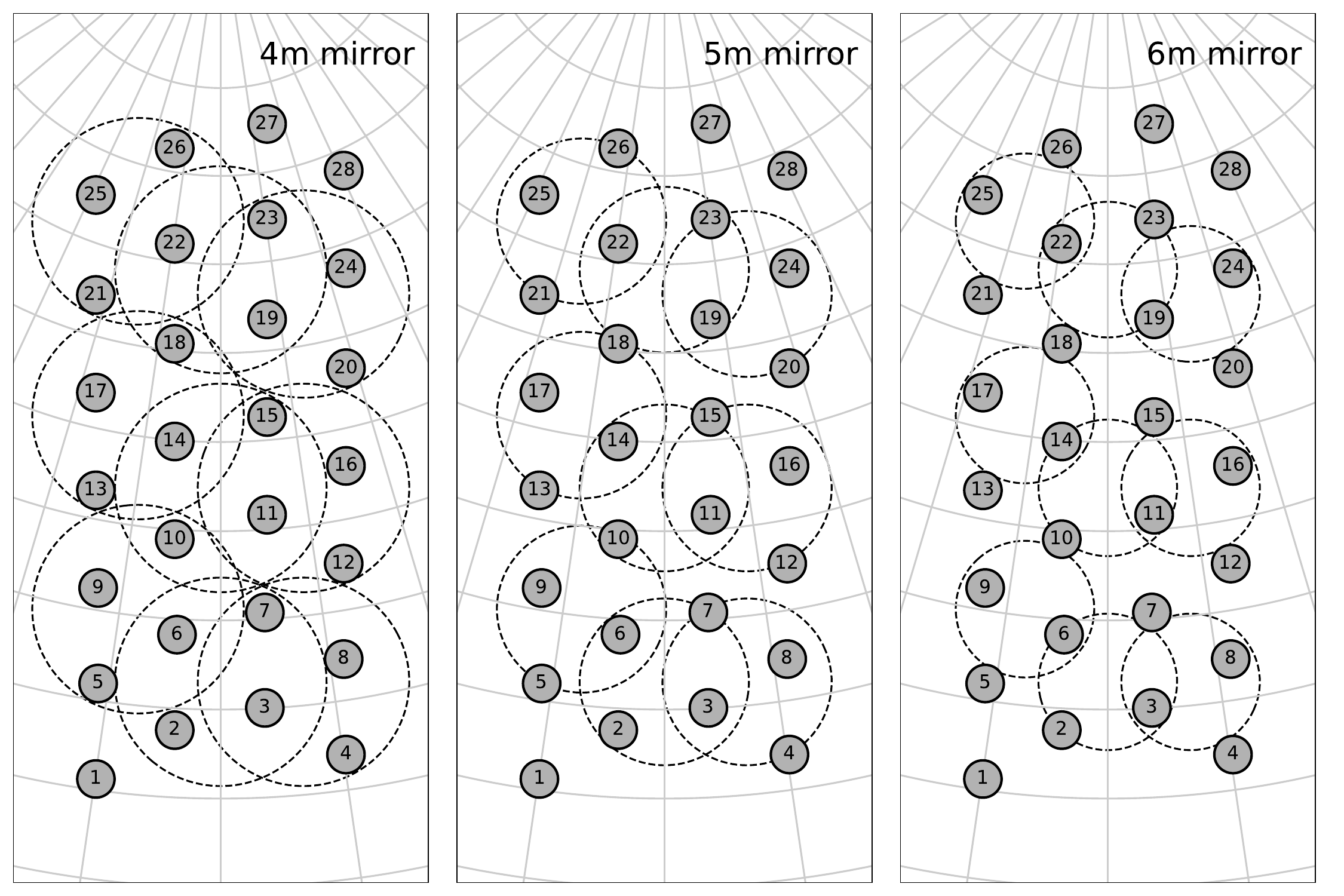}
    \caption{BINGO horns and 9 outrigger beams, for three different mirror sizes. The dashed lines show the areas restricted by the FWHM, defined in Eq. (\ref{eq:fwhm}).}
    \label{fig:beams}
\end{figure}

We modeled the BINGO telescope as having 28 independent beams, whose values of $G$, $S_{\rm min}^{(0)}$, and $\theta_{1/2}$ are shown in Table \ref{tab:bingo_beams}. For the outriggers we choose four different types: the first one is simply a BINGO naked horn (i.e. without any mirror), and the other three consist of a horn with different mirror diameters (4 m, 5 m, 6 m); the values of $G$, $S_{\rm  min}^{(0)}$, and $\theta_{1/2}$ are found in Table \ref{tab:outriggers}. Each type of outrigger is placed in four different ways, shown in Fig. \ref{fig:pointings}, totalizing sixteen configurations. In Fig. \ref{fig:beams} we show an illustrative example of the beams for 9 outriggers and the 3 different mirror sizes. For all telescopes we choose: $T_{\rm sys} = 70 $ K, $\Delta\nu = \nu_2 - \nu_1 = 280$ MHz, $K=\sqrt{2}$ and $n_p=2$.

\begin{table*}[ht!]
\small
\centering
\begin{tabular}{c cccc}
\hline
\hline
Horn & $A_{\rm eff}$ & $G$ & $S_{\rm min}^{(0)}$ & $\theta_{1/2}$ \\
& (m$^2$) & (mK/Jy) & (mJy) & (arcmin) \\
\hline
1 & 637.8 & 231.0 & 572.7 & 49.3 \\
2 & 646.3 & 234.0 & 565.2 & 49.0 \\
3 & 650.4 & 235.5 & 561.6 & 48.8 \\
4 & 641.3 & 232.2 & 569.6 & 49.2 \\
5 & 648.4 & 234.8 & 563.4 & 48.9 \\
6 & 652.3 & 236.2 & 560.0 & 48.7 \\
7 & 652.7 & 236.4 & 559.7 & 48.7 \\
8 & 648.3 & 234.8 & 563.5 & 48.9 \\
9 & 649.8 & 235.3 & 562.2 & 48.8 \\
10 & 648.8 & 235.0 & 563.0 & 48.9 \\
11 & 647.6 & 234.5 & 564.1 & 48.9 \\
12 & 647.8 & 234.6 & 563.9 & 48.9 \\
13 & 643.0 & 232.9 & 568.1 & 49.1 \\
14 & 638.3 & 231.2 & 572.2 & 49.3 \\
\hline
\end{tabular}
\quad
\begin{tabular}{c cccc}
\hline
\hline
Horn & $A_{\rm eff}$ & $G$ & $S_{\rm min}^{(0)}$ & $\theta_{1/2}$ \\
& (m$^2$) & (mK/Jy) & (mJy) & (arcmin) \\
\hline
15 & 634.1 & 229.6 & 576.0 & 49.4 \\
16 & 640.1 & 231.8 & 570.7 & 49.2 \\
17 & 626.4 & 226.9 & 583.1 & 49.7 \\
18 & 617.8 & 223.7 & 591.3 & 50.1 \\
19 & 610.4 & 221.1 & 598.4 & 50.4 \\
20 & 620.5 & 224.7 & 588.7 & 50.0 \\
21 & 602.7 & 218.3 & 606.1 & 50.7 \\
22 & 590.9 & 214.0 & 618.2 & 51.2 \\
23 & 583.2 & 211.2 & 626.3 & 51.5 \\
24 & 596.4 & 216.0 & 612.5 & 51.0 \\
25 & 571.9 & 207.1 & 638.7 & 52.1 \\
26 & 554.3 & 200.7 & 659.0 & 52.9 \\
27 & 531.6 & 192.5 & 687.1 & 54.0 \\
28 & 560.8 & 203.1 & 651.4 & 52.6 \\
\hline
\end{tabular}
\caption{BINGO beams.}\label{tab:bingo_beams}
\end{table*}

\begin{table}[h]
\small
\centering
\begin{tabular}{c cccc}
\hline
\hline
Mirror & $A_{\rm eff}$ & $G$ & $S_{\rm min}^{(0)}$ & $\theta_{1/2}$ \\
(m) & (m$^2$) & (mK/Jy) & (mJy) & (arcmin) \\
\hline
- & 1.5 & 0.5 & 251.8 & 1034.1 \\
4 & 9.8 & 3.6 & 37.1 & 396.9 \\
5 & 15.4 & 5.6 & 23.8 & 317.5 \\
6 & 22.9 & 8.3 & 15.9 & 259.9 \\
\hline
\end{tabular}
\caption{Outrigger types.}\label{tab:outriggers}
\end{table}

\subsection{Interferometry and cross correlations}\label{sec:interf}

To recover the correct position of the source, the BIS will perform  cross-correlations between pairs of telescopes. Here we define a baseline as formed by the interferometry of two telescopes, i.e BINGO main + one outrigger or one outrigger + another different outrigger.\footnote{We have implicitly assumed that there are at least two non-parallel baselines. We have already identified places near the main telescope where the outriggers might be installed, between $\sim$ 10 and 40 km in such a way that the angular precision  is $\sim$ 1-3 arcseconds.} The interferometry between BINGO main feed horns is negligible.  Assuming a perfect time delay compensation, with no taper and unity weighting function, each baseline works as an individual telescope with sensitivity given by \citep{walker1989very, thompson2017interferometry}
\begin{equation}
    S_{\rm min,\, i\times j}^{(0)} = \sqrt{\frac{S_{\rm min, \,i}^{(0)}S_{\rm min,\, j}^{(0)}}{2}}\,,
\end{equation}
while the antenna pattern is equal to
\begin{equation}\label{eq:cross_pattern}
    P_{n,\, \rm i\times j}(\mathbf{n}) = \sqrt{\frac{P_{n, \,i}(\mathbf{n})P_{n, \,j}(\mathbf{n})}{2}}\,.
\end{equation}
In a typical interferometry problem, the antenna pattern must be multiplied by a fringe term ($\propto \cos(2\pi\nu \tau)$, where $\tau$ is the time delay between measurements of two telescopes). Assuming a perfect time delay measurement we can compensate it by substituting $\cos(2\pi\nu \tau)$ by $\cos(2\pi\nu (\tau-\tau_{\rm obs}))$; this procedure is called fringe stopping and assuming that we can always set $\tau=\tau_{\rm obs}$ the cosine becomes equal to one reducing to the form in Eq.  (\ref{eq:cross_pattern}).

The total directional sensitivity for a set of $N$ telescopes is given by \citep{walker1989very}
\begin{equation}\label{eq:total_smin}
    \frac{1}{S_{\rm min}^2(\mathbf{n})} = \sum_{i=1}^N\frac{1}{S_{\rm min,\,i}^2(\mathbf{n})} + \sum_{i=1}^{N-1}\sum_{j=i+1}^N\frac{X_{ij}}{S_{\rm min,\,i\times j}^2(\mathbf{n})},
\end{equation}
where $X_{ij}=1$ if the telescopes $i$ and $j$ are physically correlated, or $X_{ij}=0$ if they are not.

\section{Generating synthetic FRBs}\label{sec:frblip}


In order to estimate the number of detections and localizations, we need to produce reliable mock catalogs. Data in the synthetic catalog contain several physical quantities randomly generated, following a probability distribution function (PDF) chosen by the user. In this section, we detail each of the physical quantities, following \cite{Luo:2018tiy,Luo:2020wfx} and present {\tt FRBlip}, a {\tt Python} code developed for this work.  Our simulations consider only non-repeaters and extragalactic FRBs.

For the single i-th FRB the observed signal-to-noise (S/N) ratio, measured by a telescope of directional sensitivity $S_{\rm min}(\mathbf{n})$, is given by
\begin{equation}\label{eq:snr_i}
    (S/N)_i = \frac{S_{\rm peak, \,i}}{S_{\rm min}(\mathbf{n}_i)},
\end{equation}where $S_{\rm peak}$ is the peak density flux, given by \citep{Lorimer:2013roa}
\begin{equation}\label{eq:Lbolspectral}
    S_{\rm{peak}}=\frac{L_{\rm{bol}}}{4\pi d_L(z)^2}\frac{(1+z)^{\alpha+1}}{\nu_{\rm{high}}'^{\alpha+1}-\nu_{\rm{low}}'^{\alpha+1}}\left(\frac{\nu_2^{\alpha+1}-\nu_1^{\alpha+1}}{\nu_2-\nu_1}\right) \,,
\end{equation}
where $d_L(z)$ is the luminosity distance, $L_{\rm bol}$ the bolometric luminosity, $\alpha$ the spectral index, $\nu_1$ and $\nu_2$ the observed frequencies. $\nu'_{\rm low}$ and $\nu'_{\rm high}$ are the lowest and highest frequencies over which the source emits, respectively, in the rest-frame of the source. This restriction on the emission frequencies implies  a range of redshift given by
\begin{equation}
    z_{\min} = \max\left[0, \frac{\nu'_{\rm low}}{\nu_2} - 1 \right]\,
\end{equation}
and
\begin{equation}
    z_{\max} = \frac{\nu'_{\rm high}}{\nu_1} - 1\,.
\end{equation}

We use the results presented in \cite{Luo:2020wfx} to generate the FRBs, where the constraints on the free parameters of the luminosity function were obtained assuming a flat spectrum with intrinsic spectral width $\Delta\nu_0 = \nu_{\rm Luo, \, high} - \nu_{\rm  Luo, \,low}=1$ GHz.  Given that the spectrum is restricted to this specific band, the corresponding luminosity is not, strictly speaking, a bolometric luminosity. Therefore, as presented in appendix \ref{ap:luminosity_luo}, the peak flux density that we will use is
\begin{equation}\label{eq:L_luo_spectral}
    S_{\rm{peak}}=\frac{L}{4\pi d_L(z)^2}\frac{1}{\nu_{\rm{Luo,\, high}}^{\alpha+1}-\nu_{\rm{Luo,\, low}}^{\alpha+1}}\left(\frac{\nu_2^{\alpha+1}-\nu_1^{\alpha+1}}{\nu_2-\nu_1}\right) \,,
\end{equation}
where   $\nu_{\rm Luo, \,high}$ and $\nu_{\rm Luo,\, low}$ are now the highest and lowest frequencies, respectively, in which the source emits as seen by the observer. While a flat spectrum was assumed in \cite{Luo:2020wfx}  because of the sample of detected FRBs around 1.4 GHz used, we assume that the same distributions are valid for a general spectral index, at least between zero and $-1.5$. The  intrinsic spectral width of 1 GHz does not contain the exact information about the frequencies $\nu_{\rm Luo, \,high}$ and $\nu_{\rm Luo, \,low}$, however, we can take values around 1.4 GHz, i.e. $\nu_{\rm Luo, \,high}=1.4$ GHz and $\nu_{\rm Luo, \,low}=400$ MHz. 
The BINGO bandwidth is located inside this frequency interval;  other intervals were investigated in the initial estimates presented in \cite{Abdalla:2021nyj}.

From Eq. (\ref{eq:total_smin}) we conclude that the total S/N is given by:
\begin{equation}
    \text{Total}[S/N] = \sqrt{\text{Auto}[S/N]^2 + \text{Intf}[S/N]^2}\,,
\end{equation}
where $\text{Auto}[S/N]$ is the total auto-correlation signal-to-noise ratio 
\begin{equation}
    \text{Auto}[S/N] = \sqrt{\sum_{i=1}^N (S/N)_i^2}\,,
\end{equation}
and $\text{Intf}[S/N]$ is the total cross-correlation signal-to-noise ratio, given by
\begin{equation}
    \text{Intf}[S/N] = \sqrt{\sum_{i=1}^{N-1}\sum_{j=i+1}^N X_{ij}(S/N)_{i\times j}^2}\,.
\end{equation}

\noindent Therefore, the intrinsic quantities which must be simulated are $z$, $L$, $\alpha$, and $\mathbf{n}$.

\subsection{Cosmological population}\label{sec:cosmic}

\subsubsection{Redshift distribution}

The FRB spatial distribution is not known yet due to the small number of measured redshifts of the associated host galaxy \citep{heintz2020host}. Some redshift distributions for FRBs have been considered over the years, e.g. a Poisson distribution $P(z)=ze^{-z}$ motivated by the distribution of gamma-ray bursts \citep{Zhou:2014yta, Yang2016extracting} or a redshift distribution following the galaxy distribution $P(z)= z^2e^{-\beta z}$ \citep{Hagstotz:2021jzu}. The alternative possibility used here   is when the source of FRBs is homogeneous in the comoving volume. The source may not be perfectly homogeneous \citep{1988ARA&A..26..509B, Luo:2018tiy}, but due to the still limited number of localized FRBs,  a  spatial distribution uniform in comoving volume can work as a first order approximation of $f_z(z)$ \citep{Luo:2018tiy, Chawla:2021igg}:
 \begin{equation}\label{eq:z_dist}
    f_z(z) \equiv \frac{\partial V}{\partial \Omega \partial z} =  \frac{c}{1+z}\frac{r^2(z)}{H(z)}\,,
 \end{equation}
 where $\partial V/(\partial \Omega \partial z)$ is the differential comoving volume per unit solid angle per unit redshift, $c$ is the speed of light, $r(z)$ is the comoving distance, and $H(z)=H_0\sqrt{\Omega_m(1+z)^3+\Omega_\Lambda}$
is the parameterized version of the first Friedmann-Lemaitre equation.
 We use the best-fit values from the Planck collaboration \citep{Planck:2018vyg}  for the matter density parameter, $\Omega_m=0.31$, dark energy density parameter, $\Omega_\Lambda=0.69$, and Hubble constant today, $H_0=67.4$ km/s/Mpc. The term $(1+z)$ takes into account the time dilation due to the cosmic expansion. The redshift is sampled according to the distribution in Eq. (\ref{eq:z_dist}) for up to the maximum value of $z_{\max}=10$. 
 
\subsubsection{Luminosity distribution}

The luminosity function of FRBs is also still not well understood and although lognormal or power-law distributions have  previously been used \citep{Caleb:2015uuk}, the  Schechter function \citep{Schechter:1976iz} opted in \cite{Luo:2018tiy, Luo:2020wfx} seems to be favored over the others \citep{Petroff:2019tty}. Thus, we assume here that it is given by
\begin{equation}\label{eq:schechter}
    \phi(L)  = \phi^\star \left( \frac{L}{L^\star} \right)^\gamma e^{-L/L^\star}\,,
\end{equation}
where  $L^\star$ is the upper cut-off luminosity,  $\phi^\star$ is a normalization constant and $\gamma$ is the power-law index. These parameters were constrained in \cite{Luo:2020wfx} using 46 FRBs: $L^*=2.9\times 10^{44} \text{erg s}^{-1}$, $\phi^*=339 \text{ Gpc}^{-3}\text{yr}^{-1}$ and $\gamma=-1.79$.

\subsubsection{Spectral index}

The flux density of FRBs depends on the frequency as $S_\nu \propto \nu^\alpha$, where the spectral index $\alpha$ can be positive or negative. In \cite{Luo:2020wfx} $\alpha=0$ was chosen inspired by the apparently flat spectrum of  FRB 121102 with 1 GHz of bandwidth \citep{Gajjar:2018bth}. \cite{2017ApJ...844..140C} reported a lack of FRB observations in the Green Bank Northern Celestial Cap survey at 350 MHz,  indicating either a flat spectrum or a spectral turnover at frequencies above 400 MHz. However, some works \citep[e.g.][]{Lorimer:2013roa} have assumed a spectral index similar to the one observed in pulsars  \citep[$\alpha =-1.4$;][]{Bates:2013ear}. Such value is very close to the result obtained by \cite{Macquart:2018rsa} using 23 FRBs ($\alpha=-1.5$).   Based on these previous  works, we choose here to use two different values for the spectral index, $\alpha =0$ and $\alpha=-1.5$. Similar values are also used in {\tt frbpoppy} in its different population setups \citep{Gardenier:2019jit}.


\subsubsection{Number of sources}

Several estimates of the all-sky rate of  observable FRBs have been made. For instance, \cite{Thornton2013} estimated a rate of $10^4$ sky$^{-1}$day$^{-1}$ above a fluence of 3 Jy ms, while CHIME recently inferred a sky rate of 820 sky$^{-1}$day$^{-1}$ above a fluence of 5 Jy ms at 600 MHz \citep{CHIMEFRB:2021srp}. \cite{Luo:2020wfx} found event rate densities of $3.5\times 10^4$ Gpc$^{-3}$ yr$^{-1}$ above a luminosity of $10^{42}$ erg s$^{-1}$, $5.0 \times 10^3$ Gpc$^{-3}$ yr$^{-1}$ above  $10^{43}$ erg s$^{-1}$ and $3.7 \times 10^2$ Gpc$^{-3}$ yr$^{-1}$ above  $10^{44}$ erg s$^{-1}$. We estimate the rate per day per sky of detectable FRBs using the following expression \citep{Luo:2020wfx}
\begin{equation}\label{eq:total_sources}
    \lambda = 4\pi \int_0^\infty \, dz \, f_z(z)\,\int_{L_0}^\infty \,d L\,\phi(L)\,,
\end{equation}
where $f_z(z)$ and $\phi(L)$ are given by Equations (\ref{eq:z_dist}) and (\ref{eq:schechter}), respectively, and $L_0$ is  the intrinsic lower cut-off of the luminosity function inferred to be $\leq 9.1 \times 10^{41}$ erg s$^{-1}$. Using Eq. (\ref{eq:total_sources}) with the values of the minimum flux density and observed pulse width for BINGO described in the next section, we estimate the total number of cosmic FRBs to be generated by {\tt FRBlip} to be $\sim 7\times 10^4$ per day of observation. In the next sections, we will describe the methodology used to estimate the detection rate for BINGO, which will be a fraction  of this cosmic population. 

\subsection{Sensitivity Maps}\label{sec:maps}

The simplest way to estimate the detection rate is 
is to follow the approach adopted by\citep{Luo:2020wfx}, also used in \cite{Abdalla:2021nyj}, through the equation\footnote{For simplicity we  have not assumed the impact on the FRB pulse due to  intra-channel smearing at high dispersion measures or scattering \citep{Petroff:2019tty, Ocker:2022xvd}. }
\begin{equation}\label{eq:detec_rate}
\lambda_\alpha(\mathbf{n}) = \int_{z_{\min}}^{z_{\max}}\, dz \, f_z(z)\int_{L_{\min}(z, \mathbf{n})}^\infty \,d L\,\phi(L)\,.
\end{equation}
The difference between Equations (\ref{eq:total_sources}) and (\ref{eq:detec_rate}) is that the former assumes an all-sky rate, while the latter is going to be calculated for the BINGO field-of-view and for redshift values bounded by the frequency range.

The minimum luminosity $L_{\min}$ in the lower limit of integration is the maximum function $\max[L_0, L_{\rm thre}]$, where $L_{\rm thre}$ depends on the spectral index $\alpha$ and the antenna pattern
\begin{equation}
L_{\rm thre}(z, \mathbf{n}) = 4\pi d_L(z)^2 \Delta\nu \left(\frac{\nu_{\rm{ Luo, high}}^{\alpha+1}-\nu_{\rm{Luo, low}}^{\alpha+1}}{\nu_2^{\alpha+1}-\nu_1^{\alpha+1}}\right) (S/N)_{\rm thre} \,S_{\rm min}(\mathbf{n})\,,
\label{eq:luminosity}
\end{equation}
where  $\Delta\nu =\nu_2 - \nu_1$, $S_{\min}$ is the telescope minimum flux density defined in eq. \ref{eq:Smin}) and $(S/N)_{\rm thre}$ indicates the minimum allowed value for the $S/N$, ensuring that we are counting only the objects rendering a minimum $S/N$ luminosity value. The detection rate per unit of time is found integrating over the telescope field-of-view,
\begin{equation}
\lambda_\alpha = \int_{S^2} \lambda_\alpha(\mathbf{n}) d\Omega\,,
\end{equation}
where this angular integration is performed using  {\tt HEALPix} \citep{Gorski:2005}, through {\tt astropy-healpix} \citep{Astropy:2018wqo}
\begin{equation}
\label{eq:senst_map}
    \lambda_\alpha = \Omega_{\rm pix} \sum_{i=1}^{n_{\rm pix}} \lambda_\alpha(\mathbf{n}_i)\,,
\end{equation}
where $n_{\rm pix}$ and $\Omega_{\rm pix}$ depend on the resolution ($n_{\rm side}$) of the {\tt HEALPix} map.
Detection rate estimates for the BINGO configurations described previously are shown in Fig. \ref{fig:rates}.
The limitation of the present approach lies in the fact that computing complex quantities as the observation rates over the baselines is very costly. Indeed, to compute the detection rate of at least two baselines, we have to compute the sensitivity map for all the possible pairs of baselines. For three baselines we must compute it for all possible combinations of three baselines and so on.

\subsection{FRBlip}

In order to compute all quantities described in previous sections, we developed \texttt{FRBlip},\footnote{\url{www.github.com/mvsantosdev/frblip}} a new {\tt Python} package  which generates mock catalogs sorting the physical quantities as random numbers, through the distributions described in Section \ref{sec:cosmic}. The information about the cosmic population is coded inside an object called \texttt{FastRadioBursts}, and the telescopes in objects of the type \texttt{RadioTelescope}. The results of observations come from the interaction between these two entities. To sort random numbers in the described distributions the code uses the \texttt{rv\_continuous} generic class, from \texttt{scipy.stats} module, to construct, by subclassing, new classes that implement each distribution. This is a high level tool that allows the developer  to implement a random number generator of any distribution.

Another important tool is the \texttt{astropy.coordinates} module, which is used to transform the non-local spherical coordinates to local coordinates at the telescope site, to perform the observations.

The dependencies of the \texttt{FRBlip} include traditional \texttt{Python} numerical libraries such as \texttt{Numpy} \citep{harris2020array}, \texttt{Scipy} \citep{2020SciPy-NMeth} and \texttt{Pandas} \citep{reback2020pandas}; high performance collection libraries as \texttt{Xarray} \citep{hoyer2017xarray} and \texttt{Sparse}; the physical numerical libraries \texttt{astropy} \citep{Astropy:2018wqo}, \texttt{HEALpix} \citep{Gorski:2005} and \texttt{Pygedm} \citep{Price:2021gzo};
and the numerical computing packages for cosmology:
\texttt{CAMB} \citep{Lewis:2002ah}, and \texttt{PyCCL} \citep{LSSTDarkEnergyScience:2018yem}.


\section{Results and discussion}\label{sec:results}

\subsection{Detecting Bursts}

We  evaluate here a more accurate detection rate. For that, we generated cosmological FRB mock catalogs using {\tt FRBlip} and counted how many of those would be detected by the main BINGO, the outriggers, and the total BIS in different scenarios. In order to  validate these results we compare them with the sensitivity map results (which we label as
``exact''). The key quantity is the yearly rate.

In order to reduce computational processing we perform a re-sampling over one day mock. The idea is that the most costly computational process is the coordinate transformation, thus, avoiding this step, reduces the processing time. To do that, we sample by first generating one day of cosmological FRBs, then we re-sample over all the variables but the sky positions (right ascension and declination), to generate one more day. We create a one year mock taking this step further 363 times. 

This procedure is then repeated 1000 times until we have enough data to adequately fit a Poisson distribution, which we did by using the \texttt{statsmodels} library \citep{seabold-proc-scipy-2010}.

\begin{figure}[h]
\centering
\includegraphics[width=\columnwidth]{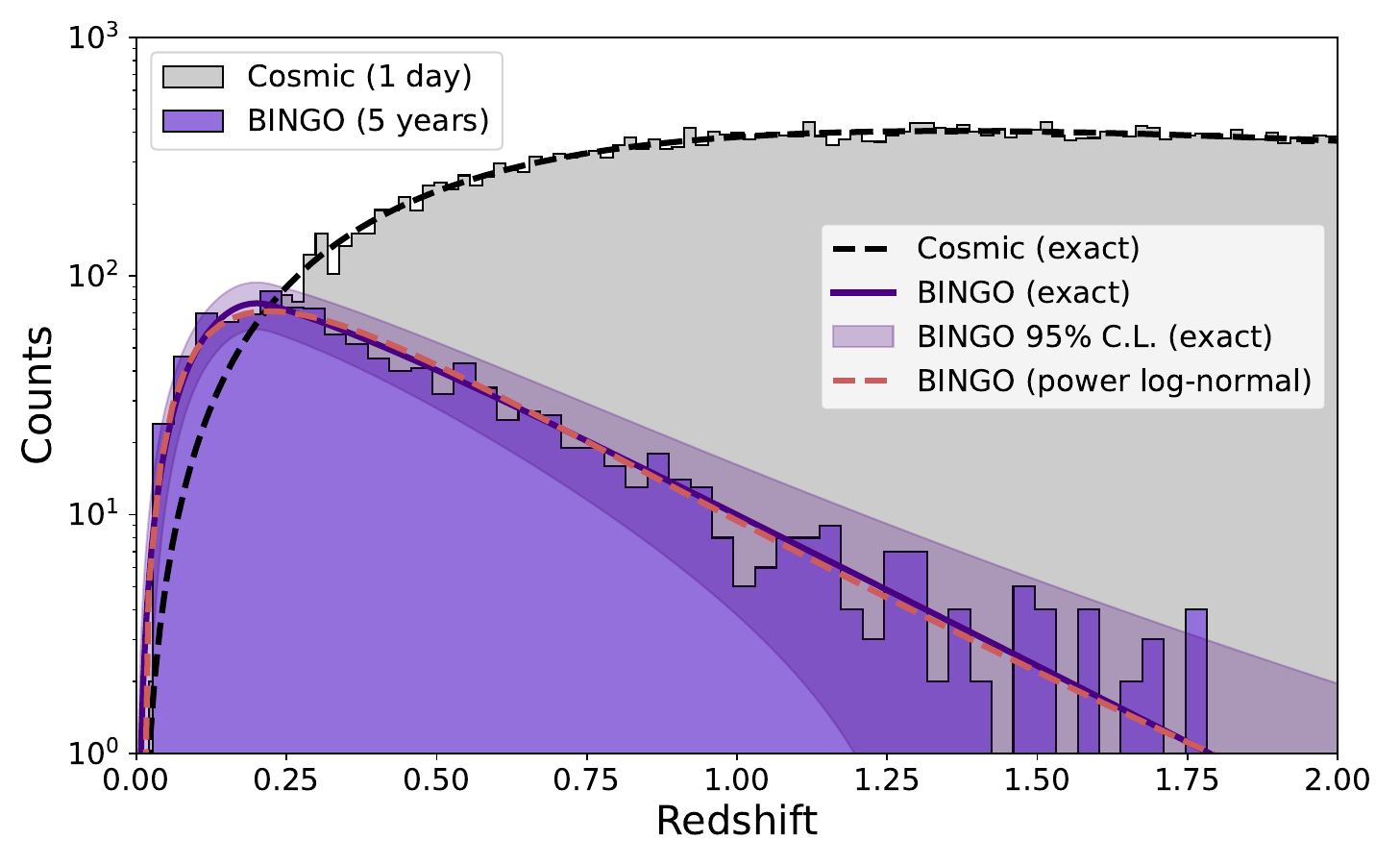}
\caption{Redshift distribution of the observed FRBs, with $S/N \geq 1$, by BINGO in five years (purple histogram) compared with all the cosmic bursts in one day (grey histogram). The exact cosmic distribution is the black dashed line, the exact distribution for the BINGO is the solid purple distribution, and the shaded region is the exact $95\%$ C.L. from the Poisson distribution. The red dashed line is the power log-normal distribution fitted from 1000 years of observation mock.}
\label{fig:redshift_distribution}
\end{figure}

We first investigate the detectability of individual telescopes. In Fig. \ref{fig:redshift_distribution} we show the redshift distribution of the FRBs seen by the main BINGO alone ($S/N \geq 1$) during five years (purple histogram) compared with all the FRBs in the sky in one day (gray histogram). This histogram is also compared with the exact distribution (purple continuous curve), computed from the sensitivity maps (\ref{eq:senst_map}).  We can see that the histogram is well bounded inside the $95\%$ confidence level (C.L.) (represented by the purple-shaded region). The red dashed line is the power log-normal\footnote{\url{docs.scipy.org/doc/scipy/reference/generated/scipy.stats.powerlognorm.html}} distribution fitted on the complete 1000 years of simulation, which is in good agreement with the exact value by $95\%$ C.L.

The number of FRBs increases with the redshift since the volume also increases until it reaches a maximum value. After that, it starts  decreasing because the luminosity limit starts  dominating. It reaches a maximum value at $z\simeq 1.8$, where the rate becomes smaller than one. 
Therefore, we can interpret $z=1.8= z_{\rm max}^{\rm eff}$, the maximum effective  redshift or the depth of the survey.

\begin{figure}[ht]
\centering
\includegraphics[width=\columnwidth]{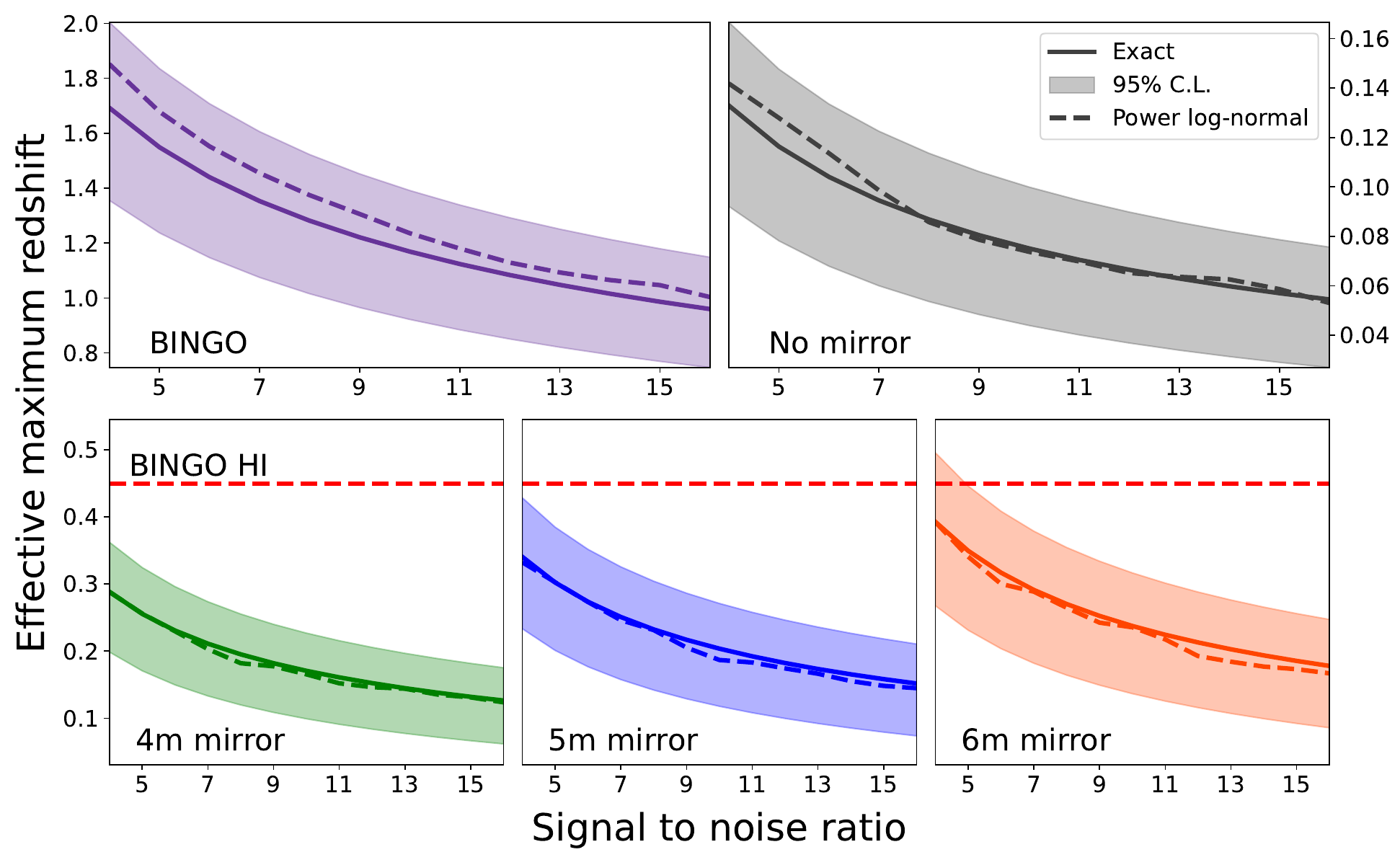}
\caption{Maximum effective redshift for different BINGO configurations, varying from $S/N \geq 5$ to $S/N \geq 15$. Comparing the exact solution from sensitivity maps (solid lines), with the fitted power log-normal distribution (dashed lines), we note that all of them are in agreement with the exact solution in $95\%$ C.L. (shaded regions). The red dashed line shows the maximum redshift of the BINGO HI survey.}
\label{fig:maximum_redshift}
\end{figure}

Fig. \ref{fig:maximum_redshift} shows how the detection rate of FRBs varies with the S/N for the different telescope configurations described in Section \ref{sec:bis}. In all cases, we see that the $z_{\rm max}^{\rm eff}$ values inferred from the power log-normal distribution, fitted from the mocks, are in agreement with the exact value by $95\%$ C.L. This alternative method to infer $z_{\rm max}^{\rm eff}$ is important to determine the depth of the survey since we can not compute the exact values from sensitivity maps as discussed in Section \ref{sec:maps}.


\begin{figure*}[ht!]
\centering
\includegraphics[width=1.8
\columnwidth]{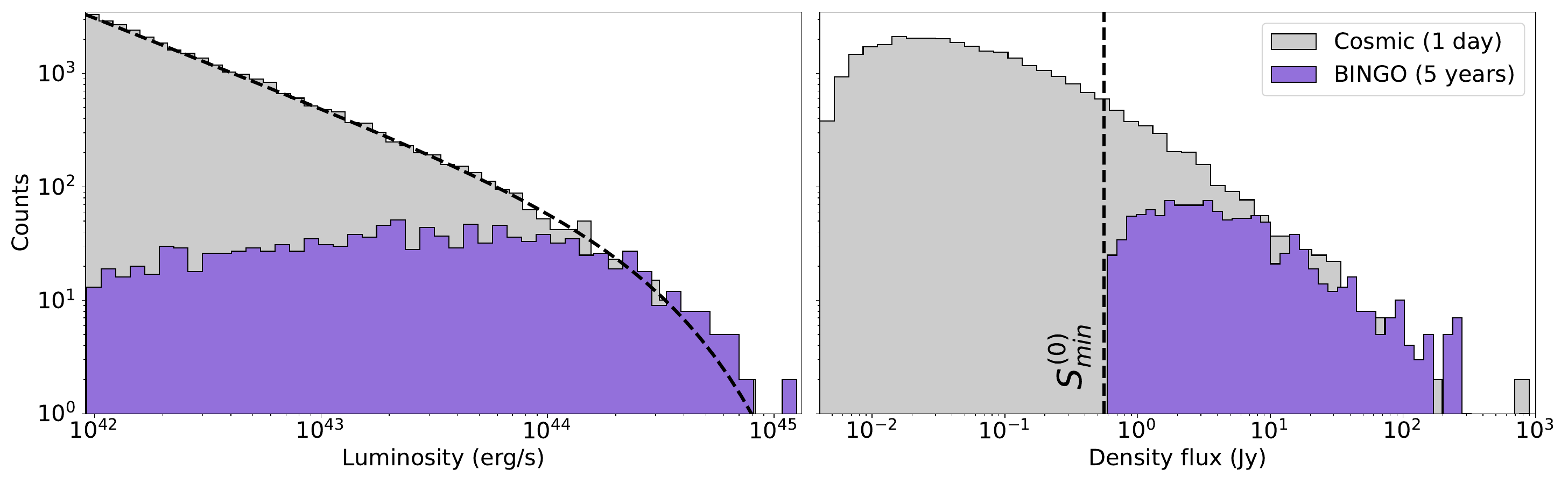}
\caption{
\textit {Left}: number of events as a function of the luminosity distribution. \textit {Right}: same, as a function of the  flux density  distribution. Estimates are for cosmic FBRs during 1 day (gray) and observed FRBs, with $S/N \geq 1$, during the 5 years of the BINGO Phase 1 mission (purple). 
}
\label{fig:luminosity_distribution}
\end{figure*}

In Fig. \ref{fig:luminosity_distribution} we show the luminosity (left) and density  flux (right) distributions. Fainter objects are more difficult to observe, as expected because it is more probable to have a density flux smaller than the minimum $S_{\rm min}^{(0)}$. In order to illustrate the fraction of the distributions observed by BINGO, we compare the histograms of the cosmic distributions in one day with the ones observed by BINGO in 5 years.  

\begin{figure}[ht]
\centering
\includegraphics[width=\columnwidth]{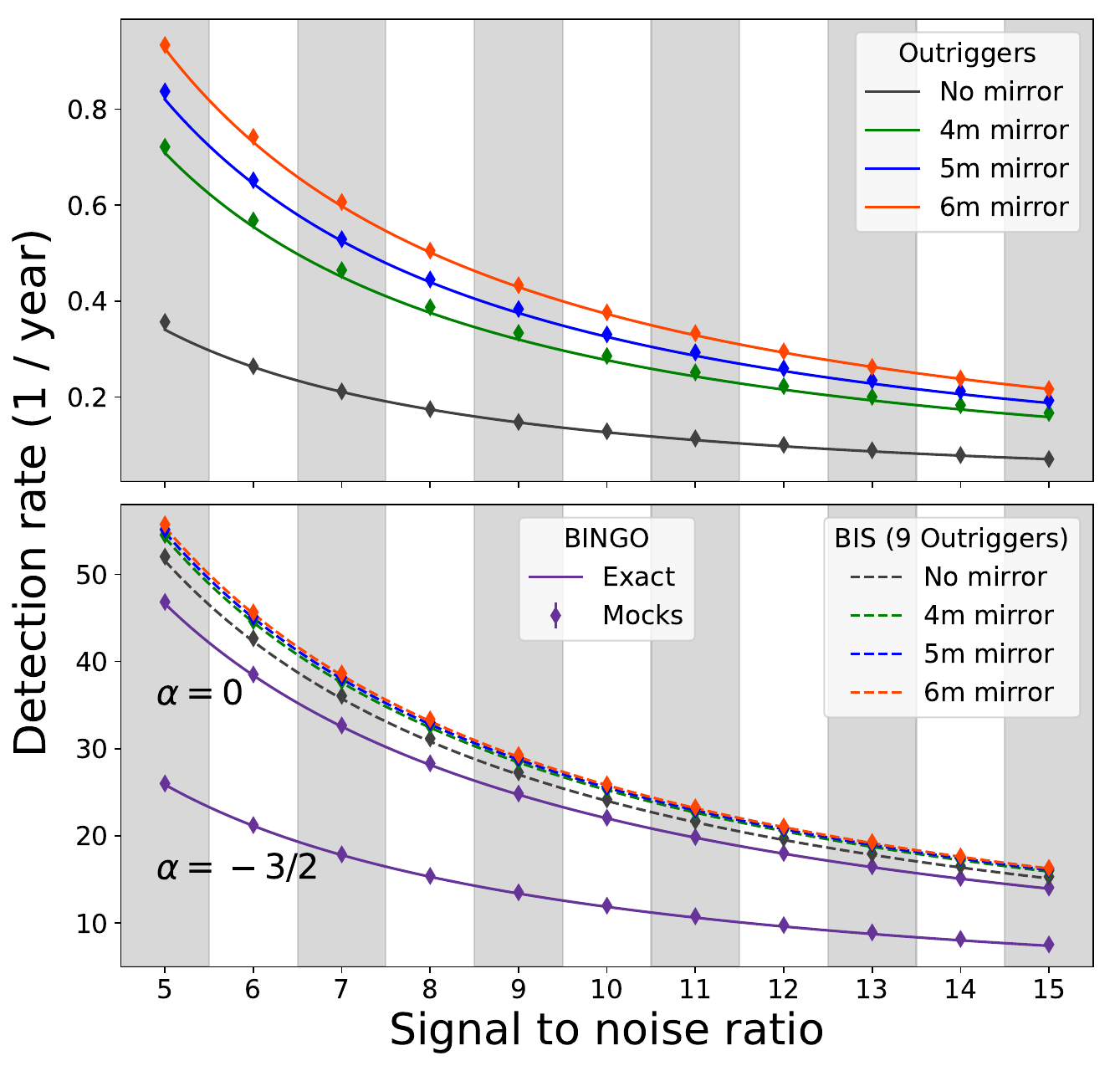}
\caption{Detection rate estimates for BINGO. \textit {Top}: mean detection rate for individual outriggers with configuration as described in Sec. \ref{sec:bis} and Table \ref{tab:outriggers}. \textit {Bottom}: detection rate for the complete BIS (main BINGO and outriggers). In both cases, we show the detection rates computed by the two methods: sensitivity maps (solid lines) and mock catalogs created with {\tt FRBlip }(scatter).}
\label{fig:rates}
\end{figure}

Finally, we show in Fig. \ref{fig:rates} the detection rates obtained from mocks and the ones  computed by the sensitivity maps, evidencing the agreement between the methods. The rates from individual outriggers (top panel) are less than one per year, however, the interferometry system, which integrates nine of these telescopes, can increase the BINGO detection rate by about $20\%$. For the case of interferometry, we have not computed the exact distributions, due to the reasons discussed in the end of Section \ref{sec:maps}. Thus, we need an alternative method to infer the depth of these interferometric cases.

\subsection{Localizing Bursts}

We now evaluate the effectiveness of BIS to localize FRBs. We assume a perfect delay compensation, i.e., the exact time delay between the telescopes that compose each baseline is known. In this case, the S/N detection of localizing an FRB increases with the number of baselines that are used to observe  it. 
Therefore, we have to select the better BIS configuration between the two options: more outriggers with narrower beam widths or fewer outriggers  with wider beam widths. Unless otherwise stated, we use a flat spectrum in this section.

The number of FRBs detected by different baselines depends on the number of outriggers used in the system and the size of the mirror. The relevant instrumental parameters for FRB detection are the sensitivity and the field
of view. Field of view generally varies inversely with the mirror size, while sensitivity increases  with increasing mirror size. Ultimately, the choice of mirror size will depend on the number of outriggers built and on the number of baselines needed to detect the same FRB. Increasing the number of outriggers also increases the collecting area and, in turn, reduces the minimum flux density.

\begin{figure}[ht]
\centering
\includegraphics[width=\columnwidth]{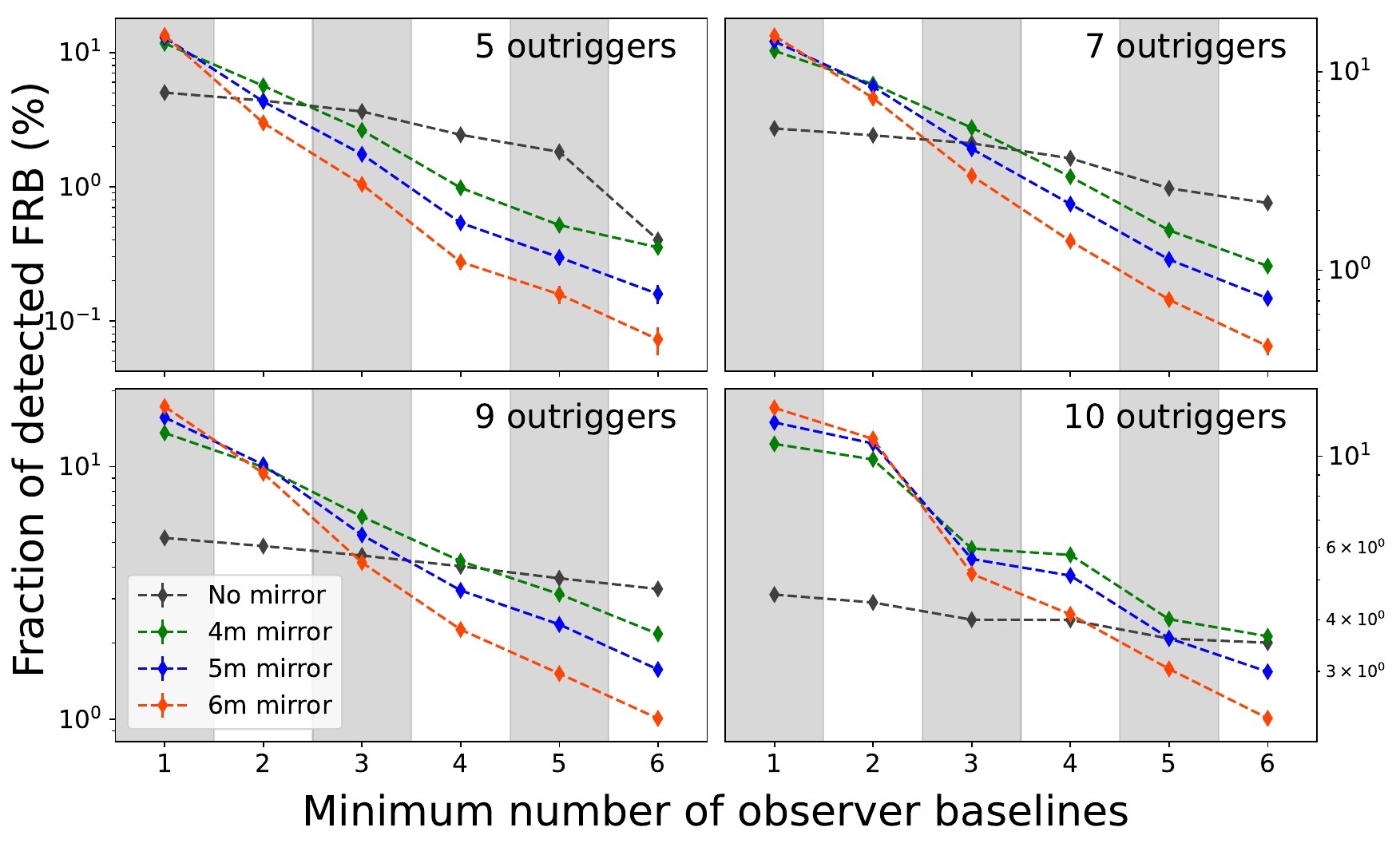}
\caption{The fraction of observed bursts ($S/N \geq 10$) as a function of baselines that detected the same events,  for the sixteen different BIS configurations.}
\label{fig:baselines_rates_2}
\end{figure}

In order to decide what is the best configuration for a BIS composed only of single-horn outriggers  we must define the following points: (i) by how many baselines an FRB must be observed to have its position well inferred; (ii) how many outriggers we can construct; and (iii) what kind of outriggers will be constructed. The answer to the first question depends only on how effective the pipeline to infer the positions is. Assuming that a minimum of two baselines is enough to pinpoint the source position, from Fig. \ref{fig:baselines_rates_2} we may infer the best type of outrigger for each configuration.

For 5 outriggers, the best choice is the 4 m mirror, for 7 outriggers the performance of 4 m and 5 m mirrors are almost the same, for 9 outriggers 4 m, 5 m, and 6 m perform approximately equal, and for 10 outriggers the 6 m mirror is the best. Now if we make it more restrictive, requiring at least 3 baselines the picture changes a bit. For 5 outriggers, for instance, now the horn with no mirror performs best, while for the other configurations (7, 9, and 10 outriggers), the 4-m mirror produces the best result.

The number of outriggers would also depend on the number of baselines needed to permit a good localization of the source. In order to understand how the number of detections per baseline affects our results, we show in Fig. \ref{fig:baselines_rates_2} the fraction of detected FRBs as a function of the number of baselines that detected a given FRB. We set the S/N to ten or higher. For this illustrative choice of S/N, we see that the fraction of detected FRBs follows the same behavior in all 4 different panels. The outriggers with a 4-m size detect more (or at least the same number) events than the ones with larger mirrors, with the exception of 10 outriggers with two baselines. However, if more baselines detect the same FRB, then the naked horn is the best option for 5 or 7 outriggers, that is, for 5 outriggers the naked horn is better if the detections are made by three baselines or more, for 7 outriggers the naked horn is better if the detections are made by four baselines or more, and so forth.

So far we have estimated the detection rates considering that the baselines and the auto-correlations have the same S/N threshold.  However, this choice is arbitrary and we may set a different approach, for instance,  to first select a group of candidates observed by the main BINGO, and then  filter this group with a different choice of S/N to define a detection in a baseline. In order to investigate how these different choices affect our results, we defined a set of values for the  S/N that will categorize  an FRB as a candidate, a detection, an interferometry detection (for the cross-correlation between main BINGO and outriggers), and localizations in one, two or three baselines. This set of definitions is shown in Table  \ref{tab:prescription}.


In Table \ref{tab:prescription}, the condition $s_1$, corresponding to the label ``Candidates'' and S/N $\ge 1$, is chosen to select the events that might be an actual FRB detection. Such S/N does not need to be much larger than 1 and can occur either in auto or cross-correlation since at this level we are simply looking for at least one baseline that received a potential signal. For the label ``Detection'' (meaning an actual detection), we choose $s_2 = 5$ in order to pick candidates that certainly would be detected by the telescope. The same occurs for $s_3$ in the interferometric detection, although in this step we already know that a detection occurred. Here we are selecting $s_3=3$ because we are interested in knowing if the FRBs would be clearly detected in interferometry side, aiming to pinpoint the source location. Furthermore, we need to avoid the situation where the interferometric detection has S/N larger than $s_3$ but no individual baseline has S/N $> 1$. We need to guarantee that at least the number of baselines we are interested in (1, 2, or 3) are measuring the signal with a consistent S/N value, and for this, we have selected $s_4=2$.

\begin{table}[h]
\footnotesize
\centering
\begin{tabular}{ c c}
\hline
\hline
Label & Condition \\
\hline
Candidate & If a given event has $S/N \geq s_1$, either \\ 
    & for auto or cross-correlations \\

Detection & Candidate with $\text{Total}[S/N] \geq s_2$ \\
Interferometric detection & Detection with $\text{Intf}[S/N] \geq s_3$ \\
\hline
\multirow[c]{3}{*}{Localization/Baselines} & Number of baselines that observed \\
    &  the same FRB satisfying $S/N \geq s_4$ \\
    & for an interferometry detection \\
\hline
\end{tabular}
\caption{$s_1 = 2$, $s_2 = 5$, $s_3 = 3$, and $s_4 = 2$.}\label{tab:prescription}
\end{table}

\begin{figure*}[ht]
\centering
\includegraphics[width=1.8
\columnwidth]{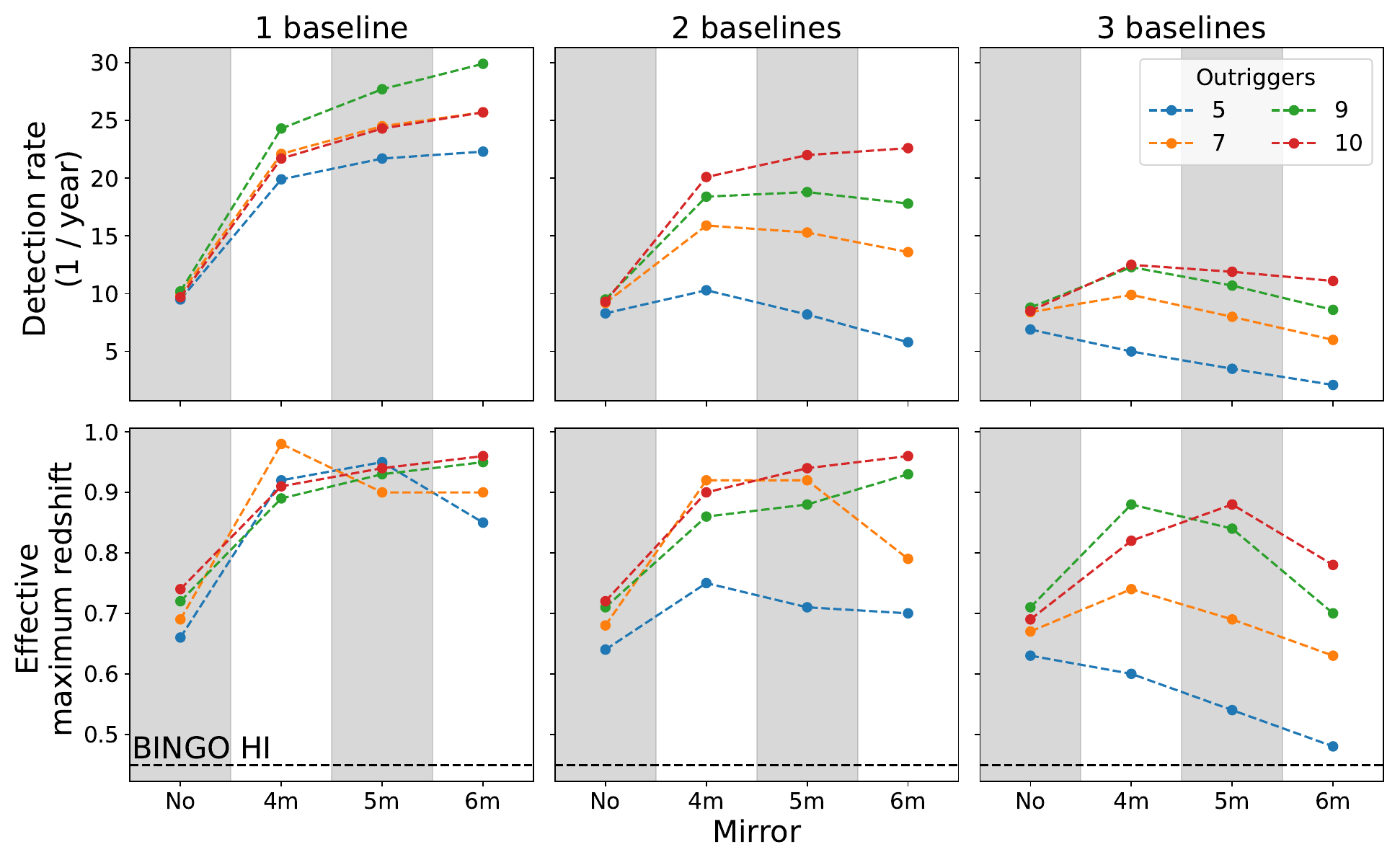}
\caption{Yearly detection rate (top) and effective maximum redshift (bottom), for prescription described in Table \ref{tab:prescription}, as a function of the mirror size for 5, 7, 9, and 10 outriggers.}
\label{fig:detec_loc}
\end{figure*}

\begin{figure*}[h]
\centering
\includegraphics[width=0.67\columnwidth]{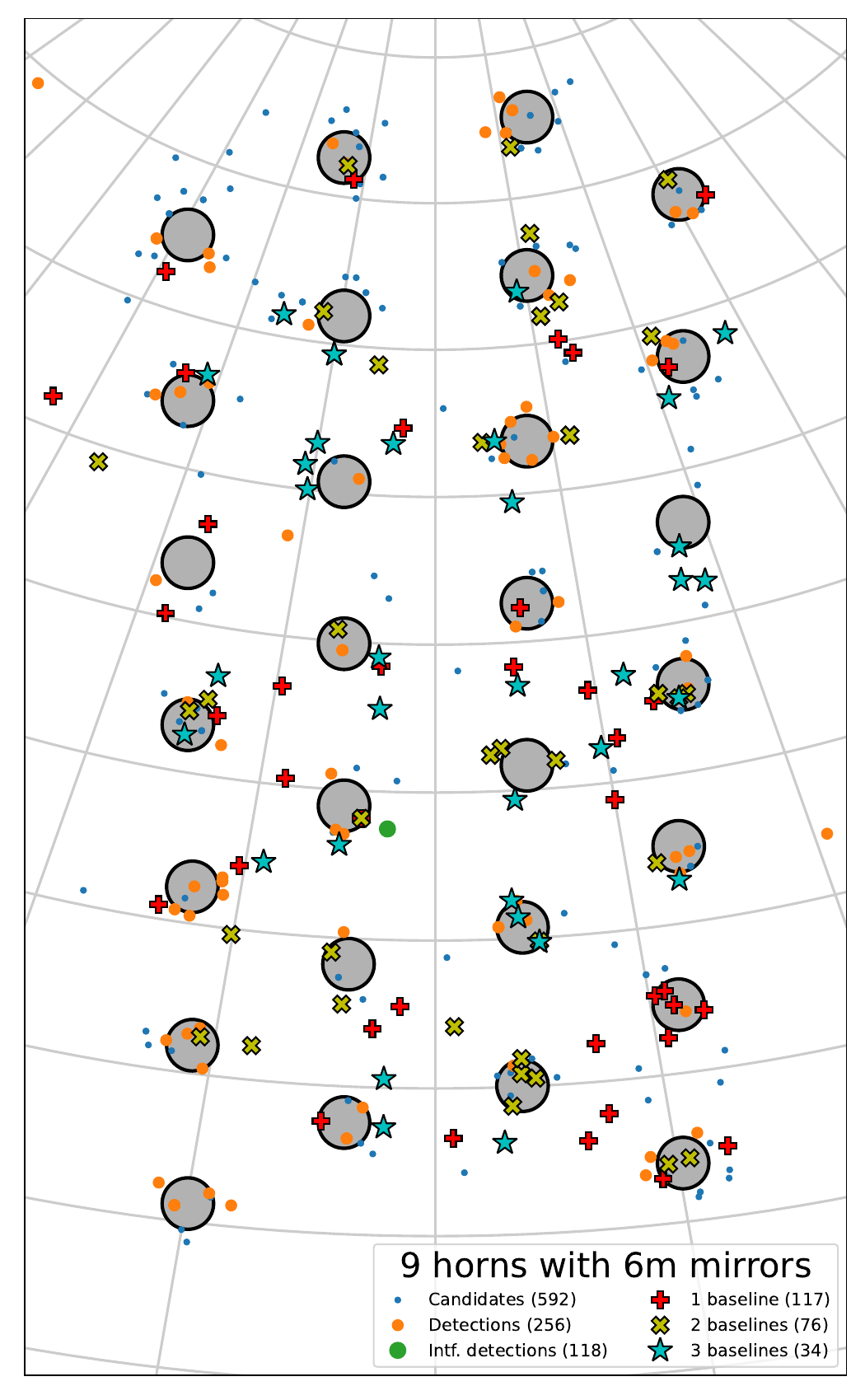}
\includegraphics[width=0.67\columnwidth]{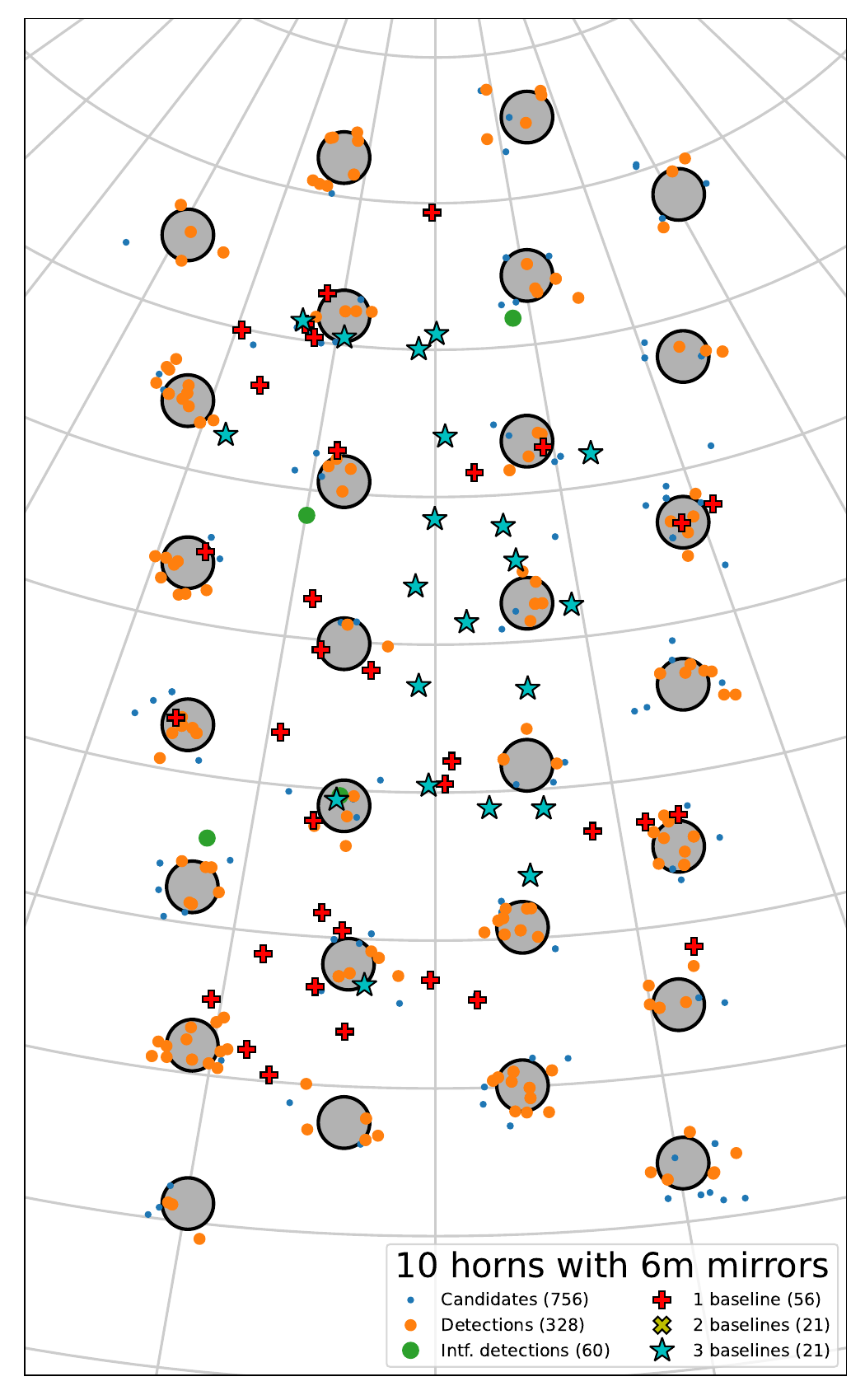}\\
\caption{ Distribution of candidates and detections for two outrigger configurations.
}
\label{fig:detec_loc_sky}
\end{figure*}

In Table \ref{tab:detec_loc_rate} and in the top panel of Fig. \ref{fig:detec_loc} we show the results for the  categorization described in Table \ref{tab:prescription}, and in Fig. \ref{fig:detec_loc_sky} we show the distribution of events in the focal plane for two outriggers setups.

The number of candidates increases slightly with the increase in the number of outriggers and mirror sizes. On the other hand, the number of detections, interferometric detections, and localizations using one baseline increase considerably with the number of outriggers and mirror sizes. The localizations using two or three baselines, however, are almost in all cases larger for a 4-m  mirror with  5, 7, or 9 outriggers. The number of localizations for a specific mirror size 
roughly increases with the number of outriggers.  However this does not always mean an increase in coverage area and the exception is the scenario with 10 outriggers, where  there are many overlaps between the outriggers' beams and main BINGO's (as can be seen from Fig. \ref{fig:pointings}), thus in practice the total observed area is less than the area of the scenario with 5 outriggers case. 


As it can be seen in Table \ref{tab:detec_loc_redshift} and bottom panel of Fig. \ref{fig:detec_loc}, the  effective maximum redshift is  between  2 and 3 for candidates and detections, but it is reduced to generally $0.5- 1$ for the localization.

We conclude from this analysis that narrow beams, i.e. horns with bigger mirrors, can observe higher redshift values but an FRB can be observed by more beams with a set composed of larger beams. This problem is compensated for by introducing more telescopes to the BIS, which expands the observed area.

As can be seen in Table \ref{tab:detec_loc_rate},  the introduction of more outriggers can multiply by many factors the detection rates. Regarding the localizations, the performance saturates for 1, 2, or 3 baselines,  reaching hundreds of localized bursts. We see similar behavior in Table \ref{tab:detec_loc_redshift} for the redshift, where the outriggers improve the depth of the survey. The result is virtually the same for 1, 2, or 3 baselines, exceeding the value of $z_{\rm max}^{\rm eff}=1$ and reaching 1.28 in the best case.





\begin{table*}
\scriptsize
\centering
\begin{tabular}{c c ccc ccc}
\hline
\hline
Number of & Mirror size & Candidates & Detections & Interferometry & \multicolumn{3}{c}{ Localizations for Baselines } \\
Outriggers & (m) & & & Detections & 1 & 2 & 3 \\
\hline
 \multirow[c]{4}{*}{5} & No dish & 121.5 & 49.5 & 11.5 & 9.5 & 8.3 & 6.9 \\
 & 4 & 124.7 & 50.9 & 16.3 & 19.9 & 10.3 & 5.0 \\
 & 5 & 125.8 & 51.3 & 16.5 & 21.7 & 8.2 & 3.5 \\
 & 6 & 126.9 & 51.7 & 16.2 & 22.3 & 5.8 & 2.1 \\
\hline
\multirow[c]{4}{*}{7} & No dish & 121.5 & 50.7 & 14.2 & 10.0 & 9.2 & 8.4 \\
 & 4 & 125.1 & 52.7 & 21.7 & 22.1 & 15.9 & 9.9 \\
 & 5 & 126.6 & 53.2 & 22.0 & 24.5 & 15.3 & 8.0 \\
 & 6 & 128.2 & 53.7 & 21.6 & 25.7 & 13.6 & 6.0 \\
 \hline
\multirow[c]{4}{*}{9} & No dish & 121.6 & 52.0 & 16.5 & 10.2 & 9.5 & 8.8 \\
 & 4 & 126.1 & 54.5 & 25.9 & 24.3 & 18.4 & 12.3 \\
 & 5 & 128.0 & 55.1 & 26.7 & 27.7 & 18.8 & 10.7 \\
 & 6 & 130.4 & 55.7 & 26.5 & 29.9 & 17.8 & 8.6 \\
 \hline
 \multirow[c]{4}{*}{10} & No dish & 119.2 & 52.3 & 17.2 & 9.7 & 9.3 & 8.5 \\
 & 4 & 121.7 & 55.3 & 26.9 & 21.7 & 20.1 & 12.5 \\
 & 5 & 122.9 & 56.0 & 28.0 & 24.3 & 22.0 & 11.9 \\
 & 6 & 124.2 & 56.6 & 28.3 & 25.7 & 22.6 & 11.1 \\
 \hline
\end{tabular}
\caption{
Observation rates (per year) for categorization are described in Table \ref{tab:prescription}. Each set of outriggers (5, 7, 9, or  10) has four options of mirrors,
The last three columns present the number of FRBs detected simultaneously (and named here as `localizations') by a different number of baselines. } 
\label{tab:detec_loc_rate}
\end{table*}

\begin{table*}
\scriptsize
\centering
\begin{tabular}{c c ccc ccc}
\hline
\hline
Number of & Mirror size & Candidates & Detections & Interferometry & \multicolumn{3}{c}{  Baselines} \\
Outriggers & (m) & & & Detections & 1 & 2 & 3 \\
\hline
 \multirow[c]{4}{*}{5} & No dish & 2.40 & 2.08 & 0.67 & 0.66 & 0.64 & 0.63 \\
 & 4 & 2.43 & 2.12 & 0.89 & 0.92 & 0.75 & 0.60 \\
 & 5 & 2.44 & 2.25 & 0.91 & 0.95 & 0.71 & 0.54 \\
 & 6 & 2.45 & 2.13 & 0.81 & 0.85 & 0.70 & 0.48 \\
\hline
\multirow[c]{4}{*}{7} & No dish & 2.40 & 2.08 & 0.73 & 0.69 & 0.68 & 0.67 \\
 & 4 & 2.43 & 2.13 & 0.98 & 0.98 & 0.92 & 0.74 \\
 & 5 & 2.44 & 2.13 & 0.88 & 0.90 & 0.92 & 0.69 \\
 & 6 & 2.45 & 2.14 & 0.88 & 0.90 & 0.79 & 0.63 \\
\hline
\multirow[c]{4}{*}{9} & No dish & 2.39 & 2.12 & 0.77 & 0.72 & 0.71 & 0.71 \\
 & 4 & 2.43 & 2.14 & 0.89 & 0.89 & 0.86 & 0.88 \\
 & 5 & 2.68 & 2.14 & 0.92 & 0.93 & 0.88 & 0.84 \\
 & 6 & 2.46 & 2.15 & 0.93 & 0.95 & 0.93 & 0.70 \\
 \hline
 \multirow[c]{4}{*}{10} & No dish & 2.37 & 2.08 & 0.79 & 0.74 & 0.72 & 0.69 \\
 & 4 & 2.40 & 2.11 & 0.93 & 0.91 & 0.90 & 0.82 \\
 & 5 & 2.40 & 2.11 & 0.95 & 0.94 & 0.94 & 0.88 \\
 & 6 & 2.41 & 2.12 & 0.99 & 0.96 & 0.96 & 0.78 \\
 \hline
\end{tabular}
\caption{
Similar to Table \ref {tab:detec_loc_rate} but now presenting the  effective maximum redshift ($z_{\rm max}^{\rm eff}$) from power log-normal distribution,  for the categorization described in Table \ref{tab:prescription}.}
\label{tab:detec_loc_redshift}
\end{table*}

\section{Conclusions and perspectives}\label{sec:conclusions}

In this work we have investigated  the capabilities of the  BINGO telescope to search and detect FRBs, and how a set of outriggers can be used to localize the events in the sky, through the BIS. We considered a single, naked horn plus three  
different mirror sizes for the outriggers, all with single horns, for 5, 7, 9, and 10 outriggers. 
In order to produce synthetic FRBs and calculate the detection rates, we developed the code {\tt FRBlip}. 

Using different methodologies to define detection and localization through baselines, we estimate that BINGO alone will be able to observe dozens of FRB per year, around 50 with $S/N \geq 5$ and 20 with $S/N \geq 15$ (for $\alpha=0$, as used in Eq. \ref{eq:luminosity})), in agreement to what was previously calculated \citep{Abdalla:2021nyj}. The introduction of outriggers can improve the total detection rate by about $20\%$ with 9 outriggers.

Regarding the localization,  if we use two baselines 
 then the best scenario is when outriggers have a 4-m mirror, and the estimates are improved from 10.3 events per year (for 5 outriggers) to 15.9 events per year (for 7 outriggers),  18.4 events per year (for 9 outriggers) or 20.1 events/year (for 10 outriggers), as seen in Table \ref{tab:detec_loc_rate}. On the other hand, if the localization is through three baselines,  the best case is for 7, 9, or 10 outriggers with a 4-m mirror, with $10-12.5$ events per year; with 5 outriggers, the best option would be outriggers without mirrors, reaching $7$ events per year. 

BINGO has the potential to increase the number of localized FRBs, such that it will be possible to better identify the host galaxy and therefore investigate inherent aspects of galaxy science related with FRBs distributions. These aspects may include  which types of galaxies allow FRB production and a possible relationship between FRB distribution and galaxy morphology and formation. Additionally, FRB detection and localization will contribute to the exploration of redshift space, which in turn can be used to better constrain cosmological parameters \citep{Walters2018}. Finally, an increased number of identified sources can help to elucidate the distribution functions of FRBs, such as the redshift distribution used in this work. 




\begin{acknowledgements}

The BINGO project is supported by São Paulo Research Foundation (FAPESP) grant 2014/07885-0. 
R.G.L. thanks Rui Luo and Mike Peel for their useful comments. 
J.Z acknowledges support from the Ministry of Science and Technology of China (grant Nos. 2020SKA0110102).  
L.S. is supported by the National Key R\&D Program of China (2020YFC2201600) and NSFC grant 12150610459. 
L.B., A.R.Q., and M.V.S. acknowledge PRONEX/CNPq/FAPESQ-PB (Grant no. 165/2018).
ARQ acknowledges FAPESQ-PB support and CNPq support under process number 310533/2022-8. 
C.A.W. acknowledges CNPq for the research grants 407446/2021-4 and 312505/2022-1.
C.P.N. thanks São Paulo Research Foundation (FAPESP) for financial support through grant 2019/06040-0. 
Y. S. is supported by grant from NSFC (Grant No. 12005184). 
X. Z. is supported by grant from NSFC (Grant No. 12005183). 
P.M. thanks São Paulo Research Foundation (FAPESP) for financial support through grant 2021/08846-2.
JRLS thanks CNPq (Grant nos. 420479/2018-0, and 309494/2021-4), and PRONEX/CNPq/FAPESQ-PB (Grant nos. 165/2018, and 0015/2019) for financial support.
This study was financed in part by the Coordenação de Aperfeiçoamento de Pessoal de Nível Superior – Brazil (CAPES) – Finance Code 88887.622333/2021-00.
The BINGO project thanks FAPESQ and the government of the State of Paraiba for funding the project. 
This research made use of {\tt astropy} \citep{Astropy:2018wqo}, {\tt healpy} \citep{Zonca2019}, {\tt numpy} \citep{harris2020array}, {\tt scipy} \citep{2020SciPy-NMeth} and {\tt matplotlib} \citep{Hunter:2007}. The authors thank the anonymous referee for the very useful report.

\end{acknowledgements}

%
%
\bibliographystyle{aa}
\bibliography{references}

\begin{appendix}
\section{Relation between bolometric luminosity and luminosity in Luo et al. (2020)} \label{ap:luminosity_luo}
The  energy released per unit of frequency interval in the rest-frame, $E_{\nu'}$, is given by \citep{Lorimer:2013roa}
\begin{equation}
   E_{\nu'}= k \nu'^{\alpha}\,,
\end{equation}
where  $k$ is a constant, $\alpha$ is the spectral index and $\nu'$ is the rest-frame frequency.
 
 The bolometric luminosity is then obtained by integrating the energy over all possible emitted frequencies
 \begin{equation}\label{eq:L_bol}
     L_{\rm bol}= \int_0^{\infty} d\nu' E_{\nu'} = \frac{k (\nu_{\rm high}'^{\alpha+1}- \nu_{\rm low}'^{\alpha+1})}{\alpha+1}\,,
 \end{equation}
 where here we have omitted the top-hat pulse of width present in \cite{Lorimer:2013roa} and $\nu'=(1+z)\nu$. 
 
 The luminosity in \cite{Luo:2020wfx}, however, is a sub-part of the bolometric luminosity, since the assumed spectral width is $\nu_{\rm Luo, \,high} - \nu_{\rm Luo, \,low}=1$ GHz.  We can then write the ``Luo'' luminosity as
 \begin{equation}\label{eq:L_luo}
     L_{\rm Luo} =   \int_0^{\infty} d\nu' B_{\nu'} E_{\nu'}= \frac{k \big(\nu_{\rm Luo,\, high}^{\alpha+1}- \nu_{\rm Luo, \,low}^{ \alpha+1}\big)}{\alpha+1}(1+z)^{\alpha+1}\,, 
 \end{equation}
where $B_{\nu'}$ is a rectangular function defined as $B_{\nu'}=1$ for $\nu_{\rm Luo, low}'\leq \nu'\leq \nu_{\rm Luo, high}'$ and $B_{\nu'}=0$ otherwise. Note that $L_{\rm Luo}= L_{\rm bol}$ if $\nu_{\rm high}'=\nu_{\rm Luo, \,high}'$ and $\nu_{\rm low}'=\nu_{\rm Luo,\, low}'$.

Using Equations (\ref{eq:L_bol}) and (\ref{eq:L_luo}) we obtain the relation between the two luminosities
\begin{equation}
    \frac{L_{\rm bol}}{\nu_{\rm high}'^{\alpha+1}- \nu_{\rm low}'^{\alpha+1}}= \frac{L_{\rm Luo}}{\big(\nu_{\rm Luo,\, high}^{\alpha+1}- \nu_{\rm Luo, \,low}^{ \alpha+1}\big)(1+z)^{\alpha+1}}
\end{equation}

Therefore, using the above relation in Eq. (\ref{eq:Lbolspectral}) we obtain Eq. (\ref{eq:L_luo_spectral}). The luminosities used in Section \ref{sec:frblip} are the ``Luo'' luminosities $L_{\rm Luo}$, where we have removed the subscript ``Luo''.

\end{appendix}

\end{document}